\documentclass[aps,preprint,showpacs,preprintnumbers,amsmath,amssymb]{revtex4}
\usepackage{amsmath,mathrsfs,amsbsy,color,graphicx,bm,amsthm,amsfonts}
\usepackage{units}
\usepackage{bbm}
\usepackage{times}
\usepackage{dcolumn}
\usepackage{mathrsfs}
\usepackage{amsmath,amssymb,epsfig}
\usepackage{amsmath}
\newcommand{\udots}{\mathinner{\mskip1mu\raise1pt\vbox{\kern7pt\hbox{.}}
\mskip2mu\raise4pt\hbox{.}\mskip2mu\raise7pt\hbox{.}\mskip1mu}}
\begin{document}
\title{Can  Hawking effect of multipartite state protect quantum resources in Schwarzschild  black hole? }
\author{Shu-Min Wu$^1$\footnote{Email: smwu@lnnu.edu.cn},  Xiao-Wei Teng$^1$, Hui-Chen Yang$^1$, Rui-Yang Xu$^1$, P. H. M. Barros$^2$\footnote{Email: phmbarros@ufpi.edu.br}, H. A. S. Costa$^2$\footnote{Email: hascosta@ufpi.edu.br} }
\affiliation{$^1$ Department of Physics, Liaoning Normal University, Dalian 116029, China\\
$^2$ Departamento de F\'{i}sica, Universidade Federal do Piau\'{i} (UFPI), Campus Min. Petr\^{o}nio Portella - Ininga, Teresina - PI, 64049-550 - Brazil}


\begin{abstract}
Most previous studies on relativistic quantum information have primarily focused on the vacuum state
$|0\rangle$ and the first excited state $|1\rangle$  in two-mode entangled systems. In this work, we go beyond these limitations by considering arbitrary $q$-th excited states  $|q\rangle$, aiming to investigate their role in preserving quantum resources. We analyze the influence of the Hawking effect on multipartite quantum states in the Schwarzschild spacetime, with particular attention to quantum entanglement and coherence. Our results show that, under the influence of the Hawking effect, increasing the excitation number $q$ leads to a reduction in quantum entanglement and mutual information, while enhancing quantum coherence. This indicates that the Hawking effect on excited multipartite states tends to degrade quantum correlations but simultaneously protects quantum coherence in curved spacetime. Therefore, when implementing quantum information protocols in gravitational settings, reducing the excitation number  $q$ is favorable for maintaining entanglement, whereas increasing $q$ may be advantageous for tasks that rely on quantum coherence in relativistic quantum information processing.
\end{abstract}

\vspace*{0.5cm}
\maketitle
\section{Introduction}
Quantum entanglement is a hallmark of multipartite quantum systems, arising from the tensor product structure of Hilbert space and the superposition principle. It plays a pivotal role as a resource in a wide range of quantum information processing protocols, including quantum cryptography, quantum dense coding, and quantum teleportation \cite{L1,L2,L3,L4,L5,L7,L8,L9,L10}. Equally fundamental is quantum coherence, which originates from the superposition of basis states and serves as a key signature of quantumness \cite{L11}. In recent years, quantum coherence has been formally recognized as a resource in its own right, with broad applications in quantum thermodynamics, condensed matter physics, biological systems, and quantum computation \cite{L12,L13,L14,L15,L16,L17}. Importantly, quantum coherence underlies the formation of entanglement and is often regarded as a necessary precursor to the generation of quantum correlations. It is well established that, through appropriate quantum operations, coherence can be converted into entanglement, highlighting a deep structural and operational connection between the two \cite{L18,L19,L20}. This foundational relationship has motivated extensive research into their interplay, resource interconversion, and role in various physical settings. Despite significant progress in exploring the interplay between coherence and entanglement, numerous open questions and challenges persist.

Hawking's seminal prediction that quantum vacuum fluctuations near a black hole event horizon give rise to spontaneous particle–antiparticle pair creation and consequently black hole evaporation is now widely known as the Hawking effect. This phenomenon underlies the emerging field of quantum information in gravitational backgrounds, which lies at the intersection of quantum information theory, quantum field theory in curved spacetime, and general relativity. In recent years, this interdisciplinary area has witnessed substantial advances in theoretical modeling \cite{L21,L22,L23,L24,L25,L26,L27,L28,L29,L30,L31,L32,L33,L34,L35,L36,L37,L38,L39,L40,L41,L42,L43,
L44,L45,L46,L47,L48,L49,L50,L51,L52,L53,L54,L55,L56,L57,L58,L59,L60,LL21}, quantum simulations \cite{L61,L62,L63,L64,L65,L66,L67,AFL1,AFL2}, and experimental schemes \cite{L68,L69,L70}, thereby deepening our understanding of relativistic quantum phenomena. A considerable body of work has been devoted to exploring how relativistic effects influence various quantum resources, including quantum steering, entanglement, discord, coherence, and mutual information \cite{L21,L22,L23,L24,L25,L26,L27,L28,L29,L30,L31,L32,L33,L34,L35,L36,L37,L38,L39,L40}. Most existing studies restrict their analysis to relatively simple quantum states, such as the maximally entangled Bell state $|\psi\rangle_{1}=\frac{1}{\sqrt{2}}(|00\rangle+|11\rangle)$, the Greenberger-Horne-Zeilinger (GHZ) state $|\psi\rangle_{2}=\frac{1}{\sqrt{2}}(|000\rangle+|111\rangle)$, and the W state $|\psi\rangle_{3}=\frac{1}{\sqrt{3}}(|001\rangle+|010\rangle+|100\rangle)$  with the quantum fields typically prepared in either the vacuum state $|0\rangle$ or the first excited state $|1\rangle$. In other words, prior investigations have predominantly focused on how the Hawking effect in these low-excitation regimes affects quantum correlations and coherence. However, the Hawking effect associated with arbitrary higher excited states $|q\rangle$ remains largely unexplored, primarily due to the significant analytical and computational complexity involved. As a result, the influence of higher excitations on the degradation or protection of quantum resources is still poorly understood. This knowledge gap forms one of the central motivations of the present work. Given the importance of safeguarding quantum resources under relativistic conditions, especially for reliable quantum information processing, we are further motivated to examine whether tuning the excitation number $q$ can serve as a viable strategy for controlling or mitigating the Hawking-induced decay of entanglement and coherence. Such an approach may provide new insights into resource optimization protocols in curved spacetime settings.

In this work, we analyze the impact of the Hawking effect on quantum entanglement, mutual information, and quantum coherence of a two-mode entangled state in Schwarzschild spacetime, with the quantum field prepared in an arbitrary $q$-th excited state. Specifically, we consider a maximally entangled initial state shared between two observers, Alice and Bob, in the asymptotically flat region,
\begin{eqnarray}\label{xwt1}
|\psi\rangle=\frac{1}{\sqrt{2}}(|q_{A}\rangle|q_{B}\rangle+|q+1_{A}\rangle|q+1_{B}\rangle).
\end{eqnarray}
We assume that Alice remains stationary at infinity, while Bob hovers near the event horizon of the black hole. Within this framework, we derive analytical expressions for the physically accessible quantum entanglement, mutual information, and coherence of the two-mode system in the presence of Hawking effect of arbitrary $q$-th excited states. We perform a systematic study of how various quantum resources behave as a function of the excitation number $q$, with particular emphasis on their sensitivity to relativistic effects induced by the Hawking effect. By judiciously selecting initial states and resource measures, we aim to identify effective strategies for enhancing relativistic quantum information protocols under curved spacetime conditions. Our results reveal that lower excitations may favor entanglement-based schemes, whereas higher excitations could enhance the robustness of coherence-based tasks. This study thus offers novel insights into the interplay between quantum resources and gravitational effects, and contributes to the broader effort of developing quantum information processing in relativistic and curved spacetime backgrounds

Recent studies on quantum information in gravitational settings, particularly in the context of black holes, have raised significant concerns regarding unitarity loss in curved spacetime. Specifically, the transition of pure states into mixed states, as influenced by gravitational effects such as Hawking radiation, remains a central challenge in the field. While the exact mechanism of unitarity loss is still under active investigation \cite{AD1}, much of the existing literature has focused on the interplay between Hawking radiation and quantum correlations like entanglement and coherence. In this work, we assume that the Hawking radiation process, though critical to the evolution of quantum states in black hole backgrounds, does not fully encapsulate the loss of unitarity at a foundational level. The initial two-qubit quantum pure state evolves into a three-qubit quantum pure state under the influence of Hawking radiation. To account for the causal disconnection between the exterior and interior regions of the black hole event horizon, we trace out the modes inside the event horizon, resulting in the transformation of the original two-qubit quantum pure state into a two-qubit quantum mixed state. Instead, our focus is on understanding how Hawking radiation impacts quantum resources, including quantum entanglement and coherence, in Schwarzschild spacetime. It should be emphasized that the question of whether unitarity is fundamentally violated in curved spacetime remains an open problem and is still under active debate \cite{AD2,AD3,AD4}.

The paper is organized as follows. In Sec.II, we derive the explicit form of an arbitrary $q$-th excited state in the vicinity of the event horizon in Schwarzschild spacetime. In Sec.III,
we investigate the behavior of quantum entanglement and mutual information under the influence of the Hawking effect associated with these excited states. Sec.IV is devoted to a detailed analysis of quantum coherence and its dependence on the excitation number $q$ in curved spacetime. Finally, the summary and potential directions for future work are provided in Sec.V.

\section{Structure of an arbitrary $q$-th excited state near event horizon of Schwarzschild  black hole}
The metric of a Schwarzschild black hole  is given by
\begin{eqnarray}\label{S1}
ds^{2}=-\bigg(1-\frac{2M}{r}\bigg)dt^{2}+\bigg(1-\frac{2M}{r}\bigg)^{-1}dr^{2}+r^{2}(d\theta^{2}+\sin^{2}\theta d\varphi^{2}),
\end{eqnarray}
where the parameter $M$ denotes the mass of the Schwarzschild black hole, and $r$ represents the radial coordinate.
For simplicity, we adopt natural units by setting $\hbar=G=c=k=1$ throughout this paper.
The dynamics of a massless scalar field in this curved background is governed by the  Klein-Gordon equation \cite{L23,L26}
\begin{eqnarray}\label{S2}
\frac{1}{\sqrt{-g}}\frac{\partial}{\partial x^{\mu}}(\sqrt{-g}g^{\mu\nu}\frac{\partial\Psi}{\partial x^{\nu}})=0.
\end{eqnarray}
The normal mode solution can be written as
\begin{eqnarray}\label{S3}
\Psi_{\omega lm}=\frac{1}{R(r)}\chi_{\omega l}(r)Y_{lm}(\theta,\varphi)e^{-i\omega t},
\end{eqnarray}
where $\chi_{\omega l}$ satisfies the radial equation
\begin{eqnarray}\label{S4}
\frac{d^{2}\chi_{\omega l}}{dr^{2}_{\ast}}+[\omega^{2}-V(r)]\chi_{\omega l}=0.
\end{eqnarray}
The corresponding mode functions are obtained by solving Eq.(\ref{S4}).
The ingoing mode, analytic across the manifold, is
\begin{eqnarray}\label{S5}
\Psi_{in,\omega lm}=e^{-i\omega \nu}Y_{lm}(\theta,\varphi),
\end{eqnarray}
while the outgoing modes, analytic only in their respective regions, read
\begin{eqnarray}\label{S6}
\Psi_{out,\omega lm}(r>r_{+})=e^{-i\omega \mu}Y_{lm}(\theta,\varphi),
\end{eqnarray}
\begin{eqnarray}\label{S7}
\Psi_{in,\omega lm}(r<r_{+})=e^{i\omega \mu}Y_{lm}(\theta,\varphi),
\end{eqnarray}
where $\nu=t+r_{\ast}$ and $\mu=t-r_{\ast}$.
The mode functions in Eqs.(\ref{S6}) and (\ref{S7}) form complete orthogonal bases in the exterior and interior regions of the event horizon, respectively. During the second quantization, the scalar field operator outside the event horizon is expanded as
\begin{eqnarray}\label{S8}
\Phi_{out}&=&\sum_{lm}\int d\omega[b_{in,\omega lm}\Psi_{in,\omega lm}(r<r_{+})
+b_{in,\omega lm}^{\dag}\Psi_{in,\omega lm}^{\ast}(r<r_{+})\notag\\
&+&b_{out,\omega lm}\Psi_{out,\omega lm}(r>r_{+})+b_{out,\omega lm}^{\dag}\Psi_{out,\omega lm}^{\ast}(r>r_{+})],
\end{eqnarray}
where $b_{in,\omega lm}$ and $b_{out,\omega lm}$
are annihilation operators associated with the interior and exterior modes, respectively. These operators satisfy the standard vacuum conditions
\begin{eqnarray}\label{S9}
b_{in,\omega lm}|0\rangle_{in}=b_{out,\omega lm}|0\rangle_{out}=0.
\end{eqnarray}

To proceed, we introduce the generalized light-like Kruskal coordinates as
\begin{eqnarray}\label{S10}
&&U=4Me^{-\frac{\mu}{4M}}, \quad V=4Me^{\frac{\nu}{4M}},  \quad \mathrm{if}\;r<r_{+};\notag\\
&&U=-4Me^{-\frac{\mu}{4M}}, \quad V=4Me^{\frac{\nu}{4M}}, \quad \mathrm{if}\;r>r_{+}.
\end{eqnarray}
The Schwarzschild spacetime can be extended to the Kruskal spacetime by introducing the Kruskal coordinates \( (U,V) \). In these coordinates, the metric takes the form
\begin{equation}
ds^2 = \frac{2 M}{r} e^{-r/2M} (-dU dV) + r^2 d\Omega^2,
\end{equation}
where \( d\Omega^2 = d\theta^2 + \sin^2\theta \, d\phi^2 \).
Using these coordinates, the outgoing Schwarzschild modes can be expressed as
\begin{eqnarray}\label{S11}
\Psi_{out,\omega lm}(r<r_{+})=e^{-4i\omega M\ln[-\frac{U}{4M}]}Y_{lm}(\theta,\varphi),
\end{eqnarray}
\begin{eqnarray}\label{S12}
\Psi_{out,\omega lm}(r>r_{+})=e^{4i\omega M\ln[\frac{U}{4M}]}Y_{lm}(\theta,\varphi).
\end{eqnarray}
Following the construction of Ruffini and Damour \cite{L71}, one defines a complete set of orthonormal outgoing modes as
\begin{eqnarray}\label{S13}
\Psi_{I,\omega lm}=e^{2\pi\omega M}\Psi_{out,\omega lm}(r>r_{+})
+e^{-2\pi\omega M}\Psi_{out,\omega lm}^{\ast}(r<r_{+}),
\end{eqnarray}
\begin{eqnarray}\label{S14}
\Psi_{II,\omega lm}=e^{-2\pi\omega M}\Psi_{out,\omega lm}^{\ast}(r>r_{+})
+e^{2\pi\omega M}\Psi_{out,\omega lm}(r<r_{+}).
\end{eqnarray}
Thus, the scalar field $\Phi_{out}$ can then be quantized in the Kruskal frame as
\begin{eqnarray}\label{S15}
\Phi_{out}&=&\sum_{lm} \int d\omega[2\sinh(4\pi\omega M)]^{-1/2}[a_{out,\omega lm}\Psi_{I,\omega lm}\notag\\
&+&a_{out,\omega lm}^{\dagger}\Psi_{I,\omega lm}^{\ast}+a_{in,\omega lm}\Psi_{II,\omega lm}+a_{in,\omega lm}^{\dagger}\Psi_{II,\omega lm}^{\ast}],
\end{eqnarray}
where  $a_{out,\omega lm}$ is the annihilation operator defined with respect to the Kruskal vacuum, satisfying
\begin{eqnarray}\label{S16}
a_{out,\omega lm}|0\rangle_{K}=0.
\end{eqnarray}
From Eqs.(\ref{S8}) and (\ref{S15}), the Bogoliubov transformations between annihilation and creation  operators in Schwarzschild and Kruskal frames can be derived as
\begin{eqnarray}\label{S17}
a_{out,\omega lm}=\frac{b_{out,\omega lm}}{\sqrt{1-e^{-\omega/T}}}-\frac{b_{in,\omega lm}^{\dag}}{\sqrt{e^{\omega/T}-1}},
\end{eqnarray}
\begin{eqnarray}\label{S18}
a_{out,\omega lm}^{\dag}=\frac{b_{out,\omega lm}^{\dag}}{\sqrt{1-e^{-\omega/T}}}
-\frac{b_{in,\omega lm}}{\sqrt{e^{\omega/T}-1}},
\end{eqnarray}
where $T=\frac{1}{8\pi M}$  denotes the Hawking temperature \cite{L26,L31}.

By normalizing the state vector, the Kruskal vacuum of the bosonic field in Schwarzschild spacetime can be written as a two-mode squeezed vacuum state, exhibiting maximal entanglement between modes inside and outside the event horizon
\begin{eqnarray}\label{S19}
|0\rangle_{K}=\sqrt{1-e^{-\omega/T}}\sum^{\infty}_{n=0} e^{-n\omega/2T}|n\rangle_{out}|n\rangle_{in}=\hat{U}|0\rangle_{out}|0\rangle_{in},
\end{eqnarray}
where $\{|n\rangle_{out}\}$ and $\{|n\rangle_{in}\}$ denote orthonormal number states localized outside and inside the event horizon, respectively \cite{L72}. The unitary operator $\hat{U}$
represents a two-mode squeezing transformation implementing the Bogoliubov transformations between the Kruskal and Schwarzschild mode bases. It is given by
\begin{eqnarray}\label{S20}
\hat{U}=\mathrm{exp}[r(a^{\dag}_{out,\omega lm}a^{\dag}_{in,\omega lm}-a_{out,\omega lm}a_{in,\omega lm})],
\end{eqnarray}
where the squeezing parameter $r$ is determined by the Hawking temperature through $\tanh r=e^{-\frac{\omega}{2T}}$.
In this work, we further an consider arbitrary $q$-th excited state $|q\rangle_{K}$ of the bosonic field in the Kruskal vacuum. By applying the transformation from Kruskal to Schwarzschild modes, $q$-th excited state $|q\rangle_{K}$  can be formally written as
\begin{eqnarray}\label{S21}
|q\rangle_{K}=\hat{U}|q\rangle_{out}|0\rangle_{in}=\frac{1}{\cosh^{q+1}r}\sum_{n=0}^{\infty}\tanh^{n}r\sqrt{\frac{(q+n)!}{q!n!}}|n+q\rangle_{out}|n\rangle_{in}.
\end{eqnarray}
Similarly, $(q+1)$-th excited state $|q+1\rangle_{K}$  becomes
\begin{eqnarray}\label{SS21}
|q+1\rangle_{K}=\hat{U}|q+1\rangle_{out}|0\rangle_{in}=\frac{1}{\cosh^{q+2}r}\sum_{n=0}^{\infty}\tanh^{n}r\sqrt{\frac{(q+1+n)!}{(q+1)!n!}}|n+q+1\rangle_{out}|n\rangle_{in}.
\end{eqnarray}
This expression captures how the Hawking effect redistributes initial excitations and generates entanglement between the accessible and inaccessible modes, even for highly excited field configurations.

\section{Influence of Hawking effect of multipartite state on quantum entanglement and mutual information in Schwarzschild spacetime }
In this section, we study the influence of Hawking effect of arbitrary $q$-th excited state on quantum entanglement and mutual information in Schwarzschild spacetime. We begin by assuming that Alice and Bob initially share a two-mode entangled state given by Eq.(\ref{xwt1}) in the asymptotically flat region of the black hole. The quantum correlations, characterized by logarithmic negativity and mutual information, are analyzed as functions of the excitation number $q$ and the Hawking temperature $T$. After the entangled state is shared, Alice remains static in the asymptotically flat region, while Bob hovers near the event horizon of the black hole.
Applying the mode transformation relations in Eqs.(\ref{S21}) and (\ref{SS21}), the initial state Eq.(\ref{xwt1}) can be reexpressed in Schwarzschild spacetime as
\begin{eqnarray}\label{s1}
|\psi_{AB}\rangle=\frac{1}{\sqrt{2}\cosh^{q+1}r}\sum_{n=0}^{\infty}\tanh^{n}r\big[
\sqrt{\frac{(q+n)!}{q!n!}}|q_{A}\rangle|n+q_{B}\rangle_{out}|n_{B}\rangle_{in}\notag\\
+\frac{1}{\cosh r}\sqrt{\frac{(q+1+n)!}{(q+1)!n!}}|q+1_{A}\rangle|n+q+1_{B}\rangle_{out}|n_{B}\rangle_{in}\big].
\end{eqnarray}
Due to the causal disconnection between the exterior and interior regions of the event horizon, the interior modes should be traced out. This leads to a mixed density operator in the exterior region, given by
\begin{eqnarray}\label{s2}
\rho_{AB_{out}}&=&\frac{1}{2\cosh^{2(q+1)}r}\sum_{n=0}^{\infty}\tanh^{2n}r\bigg[
\frac{(q+n)!}{q!n!}|q\rangle_{A}\langle q||n+q\rangle_{B,out}\langle n+q|\notag\\
&+&\frac{1}{\cosh r}\sqrt{\frac{(q+n)!(q+1+n)!}{q!(q+1)!(n!)^{2}}}\bigg(|q\rangle_{A}\langle q+1||n+q\rangle_{B,out}\langle n+q+1|\notag\\
&+&|q+1\rangle_{A}\langle q||n+q+1\rangle_{B,out}\langle n+q|\bigg)+\frac{1}{\cosh^{2}r}\frac{(q+1+n)!}{(q+1)!n!}\notag\\
&\times&|q+1\rangle_{A}\langle q+1||n+q+1\rangle_{B,out}\langle n+q+1_{B}|\bigg],
\end{eqnarray}
which can be written in matrix form as
\begin{eqnarray}\label{s3}
\rho_{AB_{out}}=\frac{1}{2\cosh^{2(q+1)}r}\sum_{n=0}^{\infty}\tanh^{2n}r\begin{pmatrix}
\frac{(q+n)!}{q!n!} & \frac{1}{\cosh r}\sqrt{\frac{(q+n)!(q+1+n)!}{q!(q+1)!(n!)^{2}}} \\
\frac{1}{\cosh r}\sqrt{\frac{(q+n)!(q+1+n)!}{q!(q+1)!(n!)^{2}}} & \frac{1}{\cosh^{2}r}\frac{(q+1+n)!}{(q+1)!n!}
\end{pmatrix}.
\end{eqnarray}

The partial transpose  criterion provides a sufficient  condition for the detection of quantum entanglement in two-mode systems \cite{L73}. Specifically, if the partial transpose of a density matrix $\rho_{AB}$ with respect to one subsystem exhibits at least one negative eigenvalue, the corresponding quantum state is necessarily entangled. In our case, we apply the partial transpose operation on Alice's subsystem and obtain the transformed density matrix as
\begin{eqnarray}\label{s4}
\rho^{T_{A}}_{AB_{out}}&=&\frac{1}{2\cosh^{2(q+1)}r}\sum_{n=0}^{\infty}\tanh^{2n}r\bigg[\tanh^{2}r\frac{(q+n+1)!}{q!(n+1)!}|q\rangle_{A}\langle q||n+1+q\rangle_{B,out}\langle n+1+q|\notag\\
&+&\frac{1}{\cosh r}\sqrt{\frac{(q+n)!(q+1+n)!}{q!(q+1)!(n!)^{2}}}\bigg(|q+1\rangle_{A}\langle q||n+q\rangle_{B,out}\langle n+q+1|\notag\\
&+&|q\rangle_{A}\langle q+1||n+q+1\rangle_{B,out}\langle n+q|\bigg)+\frac{1}{\sinh^{2}r}\frac{(q+n)!}{(q+1)!(n-1)!}\notag\\
&\times&|q+1\rangle_{A}\langle q+1_{A}||n+q\rangle_{B,out}\langle n+q|\bigg]+\frac{1}{2\cosh^{2(q+1)}r}|q\rangle_{A}\langle q| |q\rangle_{B,out}\langle q|.
\end{eqnarray}
To identify the presence of entanglement, we focus on the eigenvalues in the $(n+q,n+q+1)$ block of $\rho^{T_{A}}_{AB_{out}}$. These are given by
\begin{eqnarray}\label{s5}
\lambda_{\pm}^{T}=\frac{1}{4\cosh^{2(q+1)}r}\tanh^{2n}r\bigg[
\frac{1}{\sinh^{2}r}\frac{(q+n)!}{(q+1)!(n-1)!}+\tanh^{2}r\frac{(q+n+1)!}{q!(n+1)!}\pm\sqrt{\chi_{n}}
\bigg],
\end{eqnarray}
where $\chi_{n}=\big[\frac{(q+n)!}{q!n!}\big]^{2}\big[\frac{n^{2}}{\sinh^{4}r (q+1)^{2}}+\frac{\tanh^{4}r(q+n+1)^{2}}{(n+1)^{2}}+\frac{2(q+n+1)(n+2)}{\cosh^{2}r (q+1)(n+1)}\big]$.
To quantify the degree of entanglement, we evaluate the logarithmic negativity, a well-known entanglement monotone defined as $N(\rho)=\mathrm{log}_{2}\|\rho^{T}\|$ \cite{L73}, where $\|\rho^{T}\|$ denotes the trace norm of the partially transposed state. The analytical expression for the logarithmic negativity is thus found to be
\begin{eqnarray}\label{s7}
N(\rho_{AB_{out}})=\mathrm{Max}\Bigg[0,\log_{2}(\frac{1}{2\cosh^{2(q+1)}r}+\Lambda)\Bigg],
\end{eqnarray}
where $\Lambda=\frac{1}{2\cosh^{2(q+1)}r}\sum_{n=0}^{\infty}\tanh^{2n}r\sqrt{\chi_{n}}$.

\begin{figure}[t]
\centering
\includegraphics[width=0.6\textwidth]{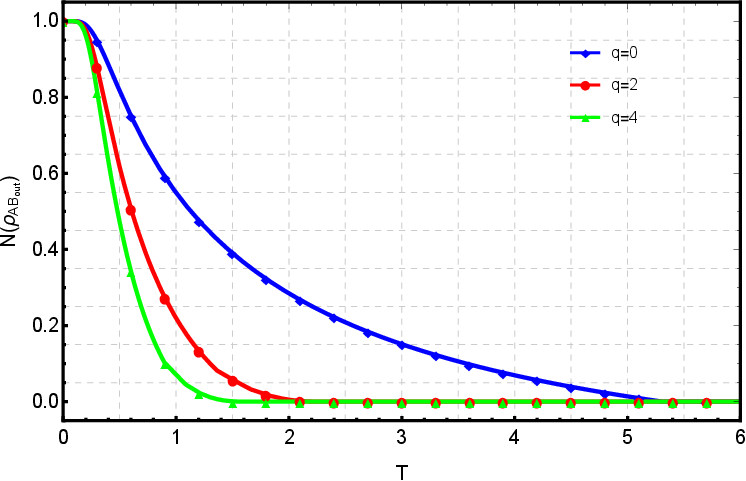}
\caption{Logarithmic negativity $N(\rho_{AB_{out}})$ as a function of the Hawking temperature $T$ for various excitation numbers $q$, with fixed frequency $\omega = 1$.}
\label{Fig1}
\end{figure}

In Fig.\ref{Fig1}, we present the logarithmic negativity $N(\rho_{AB_{\rm out}})$, which serves as a monotone quantifying the bipartite entanglement between modes $A$ and $B$, as a function of the Hawking temperature $T$ for different excitation numbers $q$, according to Eq.(\ref{s7}). For the vacuum excitation $q=0$, the logarithmic negativity decreases smoothly with increasing $T$ and asymptotically vanishes in the infinite-temperature limit. This behavior illustrates the decohering influence of Hawking radiation, which effectively acts as a thermal bath that continuously depletes quantum correlations. For nonzero excitations, such as $q=2$ and $q=4$, the decay of entanglement is markedly accelerated. This accelerated decay is due to the enhanced coupling between the initial excitations and the Hawking effect induced thermal field, which amplifies the noise and leads to a faster, irreversible loss of entanglement. Notably, the larger the initial excitation number $q$, the more rapidly the Hawking effect destroys the entanglement. These results suggest that, for preserving entanglement in relativistic quantum information protocols, it is advantageous to minimize the initial excitation number $q$.

We quantify the total correlations present in the two-mode system by evaluating the mutual information, defined as $I(\rho_{AB})=S(\rho_{A})+S(\rho_{B})-S(\rho_{AB})$, where $S(\rho)=-\mathrm{Tr}[\rho\mathrm{log}_{2}(\rho)]$ denotes the von Neumann entropy of the density matrix $\rho$ \cite{L74}. The entropy of the joint state given in Eq.(\ref{s2}) is
\begin{eqnarray}\label{s8}
S(\rho_{AB_{out}})&=&-\frac{1}{2\cosh^{2(q+1)}r}\sum_{n=0}^{\infty}\tanh^{2n}r\Bigg[\frac{(q+n)!}{q!n!}+\frac{1}{\cosh^{2}r}\frac{(q+n+1)!}{(q+1)!n!}\Bigg]\notag\\
&\times&\mathrm{log}_{2}\Bigg\{\frac{\tanh^{2n}r}{2\cosh^{2(q+1)}r}\Bigg[\frac{(q+n)!}{q!n!}+\frac{1}{\cosh^{2}r}\frac{(q+n+1)!}{(q+1)!n!}\Bigg]\Bigg\}.
\end{eqnarray}
By tracing out Alice's degrees of freedom, we obtain the reduced density matrix for Bob's accessible region, $\rho_{B_{out}}$, as given by
\begin{eqnarray}\label{s9}
\rho_{B_{out}}&=&\frac{1}{2\cosh^{2(q+1)}r}\sum_{n=0}^{\infty}\tanh^{2n}r\Bigg[\frac{(q+n)!}{q!n!}|n+q\rangle_{B,out}\langle n+q|+\frac{1}{\cosh^{2}r}\frac{(q+1+n)!}{(q+1)!n!}\notag\\
&\times&|n+q+1\rangle_{B,out}\langle n+q+1|\Bigg]=\frac{1}{2\cosh^{2(q+1)}r}\sum_{n=0}^{\infty}\tanh^{2n}r\Bigg[\frac{(q+n)!}{q!n!} \notag\\
&+&\frac{1}{\sinh^{2}r}\frac{(q+n)!}{(q+1)!(n-1)!}\Bigg]|n+q\rangle_{B,out}\langle n+q|.
\end{eqnarray}
The corresponding von Neumann entropy is given by
\begin{eqnarray}\label{s10}
S(\rho_{B_{out}})&=&-\frac{1}{2\cosh^{2(q+1)}r}\sum_{n=0}^{\infty}\tanh^{2n}r\Bigg[\frac{(q+n)!}{q!n!}+\frac{1}{\sinh^{2}r}\frac{(q+n)!}{(q+1)!(n-1)!}\Bigg]\notag\\
&\times&\mathrm{log}_{2}\Bigg\{\frac{\tanh^{2n}r}{2\cosh^{2(q+1)}r}\Bigg[\frac{(q+n)!}{q!n!}+\frac{1}{\sinh^{2}r}\frac{(q+n)!}{(q+1)!(n-1)!}\Bigg]\Bigg\}.
\end{eqnarray}
Conversely, tracing over Bob’s subsystem yields Alice's reduced density matrix
\begin{eqnarray}\label{s11}
\rho_{A}&=&\frac{1}{2}(|q\rangle_{A}\langle q|+|q+1\rangle_{A}\langle q+1|),
\end{eqnarray}
which yields a constant entropy $S(\rho_{A})=1$.
Combining these results, we arrive at the analytical expression for the mutual information in Schwarzschild spacetime
\begin{eqnarray}\label{s12}
I(\rho_{AB_{out}})&=&1-\frac{1}{2\cosh^{2(q+1)}r}\sum_{n=0}^{\infty}\tanh^{2n}r\Bigg[\frac{(q+n)!}{q!n!}+\frac{1}{\sinh^{2}r}\frac{(q+n)!}{(q+1)!(n-1)!}\Bigg]\notag\\
&\times&\mathrm{log}_{2}\Bigg\{\frac{\tanh^{2n}r}{2\cosh^{2(q+1)}r}\Bigg[\frac{(q+n)!}{q!n!}+\frac{1}{\sinh^{2}r}\frac{(q+n)!}{(q+1)!(n-1)!}\Bigg]\Bigg\}\notag\\
&+&\frac{1}{2\cosh^{2(q+1)}r}\sum_{n=0}^{\infty}\tanh^{2n}r\Bigg[\frac{(q+n)!}{q!n!}+\frac{1}{\cosh^{2}r}\frac{(q+n+1)!}{(q+1)!n!}\Bigg]\notag\\
&\times&\mathrm{log}_{2}\Bigg\{\frac{\tanh^{2n}r}{2\cosh^{2(q+1)}r}\Bigg[\frac{(q+n)!}{q!n!}+\frac{1}{\cosh^{2}r}\frac{(q+n+1)!}{(q+1)!n!}\Bigg]\Bigg\}.
\end{eqnarray}
This expression reveals how the total correlations depend explicitly on the excitation number $q$ and the Hawking temperature parameter $r$, thereby quantifying the redistribution and degradation of correlations induced by the Hawking effect.

In the high-temperature limit (\( T \to \infty \)), we find that all quantum correlations vanish. This behavior is strikingly similar to the expected results from the Page curve, which describes the entanglement entropy of Hawking radiation. According to the Page curve, entanglement entropy first increases and then returns to zero during the black hole evaporation process, assuming unitary evolution. Our findings are consistent with this expectation, as the von Neumann entropy \( S(\rho_{AB_{out}}) \) between Alice and Bob first increases from zero, reaches a maximum, and then returns to zero as \( T \) increases. Furthermore, although quantum correlations are lost at high temperatures, classical correlations persist, governed by the local entropy of Alice's subsystem. This suggests that even in the high-temperature regime, where quantum information appears to be lost, there may still be a possibility for information recovery via classical correlations. This observation offers a potential connection to the black hole information paradox, indicating that while quantum information may not be fully recovered, classical information could still provide a route for information retrieval in the black hole’s final stages of evaporation.

\begin{figure}[t]
\centering
\includegraphics[width=0.6\textwidth]{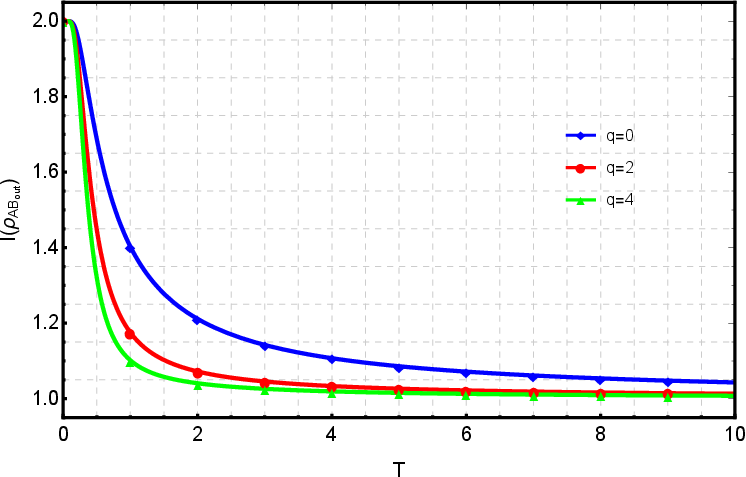}
\caption{Mutual information $I(\rho_{AB_{\text{out}}})$ as a function of the Hawking temperature $T$ for different excitation numbers $q$, with fixed frequency $\omega=1$.}
\label{Fig2}
\end{figure}

In Fig.\ref{Fig2}, the mutual information $I(\rho_{AB_{\rm out}})$ between modes $A$ and $B$ is shown as a function of the Hawking temperature $T$ for different excitation numbers $q$, as obtained from Eq.(\ref{s12}). The results indicate that $I(\rho_{AB_{\rm out}})$ decreases rapidly with $T$ and asymptotically approaches a constant value $1$ in the high-temperature regime. Similar to the behavior of quantum entanglement, mutual information is further suppressed with increasing $q$, reflecting the fact that higher excitation levels, associated with stronger multipartite participation, enhance the decoherence induced by Hawking radiation. This trend highlights the increased fragility of quantum correlations in the presence of thermal noise as $q$ grows.
Therefore, minimizing the initial excitation number is advantageous for preserving total correlations in relativistic quantum information tasks near the black hole. In the extreme limit, we obtain
$$\lim_{T\rightarrow\infty}S(\rho_{AB_{out}})=\lim_{T\rightarrow\infty}S(\rho_{B_{out}})=0,$$
which leads to
$$\lim_{T\rightarrow\infty}I(\rho_{AB_{out}})=S(\rho_{A}),$$
independent of $q$. This confirms that, at sufficiently high temperatures, the state $\rho_{AB_{out}}$ retains only classical correlations, with all quantum correlations irreversibly lost.

\section{Influence of Hawking effect of multipartite state on quantum coherence in Schwarzschild spacetime  }

In this section, we explore how the Hawking effect of arbitrary $q$-th excited state $|q\rangle_{K}$ impacts the quantum coherence  in Schwarzschild spacetime. To capture the coherence properties more comprehensively, we employ two widely used quantifiers: the $l_{1}$-norm of coherence and relative entropy of coherence (REC) \cite{L75}. The $l_{1}$-norm of coherence, an intuitive and basis-dependent measure, is defined as the sum of the absolute values of all off-diagonal elements in the density matrix $\rho$
\begin{eqnarray}\label{ss12}
C_{l_{1}}(\rho)=\sum_{i\neq j}|\rho_{i,j}|.
\end{eqnarray}
The REC, on the other hand, quantifies the distinguishability between $\rho$ and its diagonal counterpart $\rho_{\text{diag}}$ in the reference basis and is given by
\begin{eqnarray}\label{ss13}
C_{RE}(\rho)=S(\rho_{\text{diag}})-S(\rho),
\end{eqnarray}
where $S(\rho)$ denotes the von Neumann entropy of $\rho$, and $\rho_{\text{diag}}$ is obtained by removing all off-diagonal elements from $\rho$. Based on Eqs.(\ref{ss12}) and(\ref{ss13}), we derive the analytical expressions of the $l_{1}$-norm of coherence and REC for the reduced two-mode system $\rho_{AB_{\text{out}}}$ in the presence of Hawking effect as
\begin{eqnarray}\label{ss14}
C_{l_{1}}(\rho_{AB_{out}})&=&\frac{1}{\cosh^{2q+3}r}\sum_{n=0}^{\infty}\tanh^{2n}r\sqrt{\frac{(q+n)!(q+1+n)!}{q!(q+1)!(n!)^{2}}},
\end{eqnarray}

\begin{eqnarray}\label{ss15}
C_{REC}(\rho_{AB_{out}})&=&-\frac{1}{2\cosh^{2(q+1)}r}\sum_{n=0}^{\infty}\tanh^{2n}r\Bigg\{\frac{(q+n)!}{q!n!}\log_{2}\Bigg[\frac{\tanh^{2n}r}{2\cosh^{2(q+1)}r}\frac{(q+n)!}{q!n!}\Bigg]\notag\\
&+&\frac{1}{\cosh^{2}r}\frac{(q+1+n)!}{(q+1)!n!}\log_{2}\Bigg[\frac{\tanh^{2n}r}{2\cosh^{2(q+2)}r}\frac{(q+1+n)!}{(q+1)!n!}\Bigg]-\Bigg[\frac{(q+n)!}{q!n!}+\frac{1}{\cosh^{2}r}\notag\\
&\times&\frac{(q+1+n)!}{(q+1)!n!}\Bigg]\log_{2}\Bigg[\frac{\tanh^{2n}r}{2\cosh^{2(q+1)}r}\bigg(\frac{(q+n)!}{q!n!}+\frac{1}{\cosh^{2}r}\frac{(q+1+n)!}{(q+1)!n!}\bigg)\Bigg]  \Bigg\}.
\end{eqnarray}
These two measures together provide complementary perspectives: while the $l_{1}$-norm  directly reflects the magnitude of off-diagonal coherence, the REC captures its informational content by quantifying the entropy difference between the state and its decohered version.

\begin{figure}
\centering
\begin{minipage}[t]{0.48\linewidth}
\centering
\includegraphics[width=\linewidth]{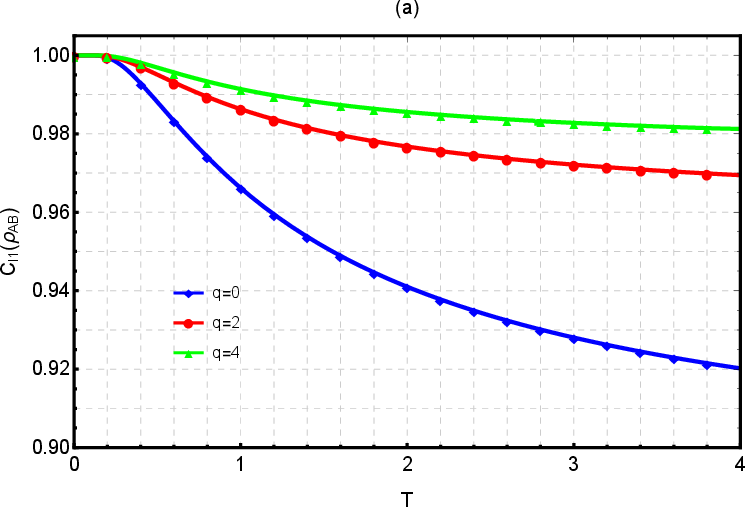}
\label{fig3a}
\end{minipage}
\hfill
\begin{minipage}[t]{0.48\linewidth}
\centering
\includegraphics[width=\linewidth]{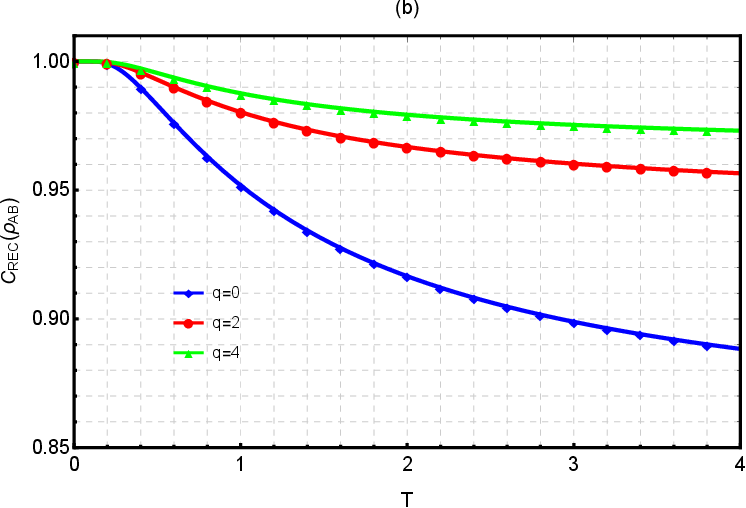}
\label{fig3b}
\end{minipage}
\caption{(a) $l_{1}$-norm of coherence and (b) REC for the system $\rho_{AB}$ as a function of the Hawking temperature $T$ for various excitation numbers $q$, with fixed frequency $\omega = 1$.}
\label{Fig3}
\end{figure}

In Fig.\ref{Fig3}, we present the dependence of quantum coherence for the system $\rho_{AB_{out}}$ on the Hawking temperature $T$ for various excitation numbers $q$. As shown, quantum coherence decreases monotonically with increasing $T$, indicating that thermal noise introduced by the Hawking effect degrades coherence. In contrast to quantum entanglement, quantum coherence asymptotically approaches a nonzero value. This demonstrates the enhanced robustness of quantum coherence relative to entanglement in the presence of Hawking radiation. Moreover, increasing the initial excitation number $q$ leads to stronger quantum coherence, whereas it typically suppresses entanglement. This inverse behavior suggests that higher $q$ values, which correspond to more complex multipartite structures, can effectively protect coherence from thermal decoherence. These findings provide valuable insights into the behavior of quantum coherence in curved spacetime and support the feasibility of quantum information protocols in extreme gravitational environments. For specific applications, if a task primarily exploits quantum coherence, a larger excitation number $q$ should be preferred. Conversely, when entanglement is the central resource, a smaller $q$ is more advantageous. Therefore, appropriate selection of the initial excitation number is crucial for optimizing the performance of quantum information processing in relativistic and high-temperature regimes.

The central result of this paper-that higher excitation number \( q \) accelerates the loss of entanglement and mutual information while enhancing quantum coherence can be understood through an intuitive physical argument. As the excitation number \( q \) increases, the quantum state becomes more complex. This leads to a stronger coupling between the quantum system and the thermal bath associated with Hawking radiation. The interaction between the system and the thermal bath, introduced through the Bogoliubov transformation, accelerates decoherence, resulting in the rapid degradation of entanglement and mutual information. The higher \( q \) is, the more modes are involved, amplifying the coupling to the Hawking radiation and increasing the thermal noise, which accelerates the loss of quantum correlations. The Bogoliubov transformation plays a crucial role in this process by mixing the modes of the quantum field inside and outside the event horizon. This mixing redistributes the excitations, and as \( q \) increases, more modes contribute to the state, leading to a stronger coupling between the system and the thermal bath. The larger number of excitations introduces a more significant interaction between the initial state and the thermal fluctuations, accelerating the decoherence process. This stronger coupling with the thermal bath explains why entanglement and mutual information are more rapidly degraded for higher excitations.

The mode-mixing structure of the Schwarzschild vacuum  affects higher excitations more severely in terms of entanglement because entanglement is sensitive to the correlations between the subsystems-specifically, between the modes outside the black hole and the observer. The increased number of excitations \( q \) results in a larger overlap between the modes inside and outside the event horizon, which leads to a faster degradation of the entanglement. This accelerated loss occurs due to the stronger coupling to the thermal field as higher excitations introduce more modes and increase the complexity of the state.
On the other hand, quantum coherence behaves differently under higher excitations. While entanglement is highly sensitive to the mode mixing and the coupling with the thermal bath, coherence is more resilient. This is because coherence is related to the internal structure of the quantum state within a single subsystem, not directly to the correlations between subsystems. As \( q \) increases, the internal complexity of the state increases, which helps shield the coherence from thermal noise. The greater the excitation number, the more superpositions are present within the system, which enhances its robustness against decoherence caused by the Hawking radiation. This explains why coherence increases with \( q \), even as entanglement and mutual information decay. Thus, increasing \( q \) modifies the coupling between the initial state and the thermal Hawking bath through the Bogoliubov transformation, which becomes stronger as more modes are involved. The mode-mixing structure of the Schwarzschild vacuum leads to a faster loss of entanglement for higher \( q \), while the increased complexity of the state helps protect quantum coherence from the thermal noise.

\section{Conclusions}
In this work, we have investigated the degradation of quantum resources, including entanglement, mutual information, and quantum coherence, due to the Hawking effect acting on multipartite bosonic excitations in Schwarzschild spacetime. By considering arbitrary $q$-excited initial states, we have systematically analyzed how the excitation number $q$ modulates the robustness of each quantum resource as the Hawking temperature increases. Our analysis shows that logarithmic negativity, a measure of entanglement, decreases monotonically with Hawking temperature, with larger $q$ leading to a faster and more pronounced loss of entanglement. This reflects the fragility of distillable entanglement under thermal noise, particularly when more excitations enhance the system's coupling to Hawking effect. Similarly, mutual information, which quantifies the total correlations, also diminishes with Hawking temperature, and the suppression becomes stronger as $q$ increases. Nevertheless, mutual information asymptotically saturates to a classical limit determined by the local entropy of Alice's subsystem, showing that while quantum correlations vanish, a residual amount of classical correlation can survive even in the extreme temperature limit.

Furthermore, we  show that quantum coherence, quantified via the $l_1$-norm  of coherence and REC, decays with Hawking temperature but remains finite in the high temperature limit.  Remarkably, quantum coherence increases with excitation number $q$, indicating that multipartite excitations introduce structural complexity that enhances the resilience of coherent superpositions to gravitational decoherence. This observation underscores a fundamental disparity in how different quantum resources behave under relativistic effects. Our results highlight that different quantum resources respond unequally to Hawking effect of the black hole, and that the choice of initial excitation number $q$ plays a dual role: larger $q$ favors coherence-based protocols, whereas smaller $q$ helps preserve entanglement. These findings offer theoretical guidance for optimizing relativistic quantum information tasks near black hole, and may also shed light on how information survives or degrades in curved spacetime.

It is also useful to interpret our results from the perspective of entanglement monogamy and multipartite correlations. From the viewpoint of the observer Alice located in the asymptotically flat region, the total correlations are not destroyed by the Hawking effect but redistributed among accessible and inaccessible modes. The initial mutual information shared between Alice and Bob is conserved at the global level: the decrease of mutual information between Alice and the exterior mode of Bob is compensated by the increase of mutual information between Alice’s mode and the inaccessible interior modes. In this sense, the physically accessible correlations are converted into inaccessible correlations rather than being fundamentally lost. This redistribution becomes more significant as the excitation number \( q \) increases, since higher excitations enhance the mode mixing between interior and exterior regions. From the perspective of multipartite entanglement \cite{LAD1}, before tracing out the interior modes the system should be regarded as a tripartite system consisting of Alice’s mode, the exterior mode, and the interior mode. In curved spacetime, the entanglement between Alice’s mode and the total system of the other modes remains equal to the initial entanglement, while the Hawking effect mainly redistributes this entanglement between the exterior and interior modes. In particular, the entanglement involving the interior modes can be generated by the Hawking effect through the Bogoliubov mode mixing. Therefore, in the high-temperature limit where the exterior bipartite entanglement vanishes, the entanglement is not completely destroyed but effectively transferred to correlations involving the inaccessible interior modes. This interpretation is consistent with the monogamy of entanglement, where the reduction of bipartite entanglement in the accessible subsystem is accompanied by the redistribution of entanglement into multipartite correlations involving the interior modes.

\begin{acknowledgments}
This work is supported by the National Natural
Science Foundation of China (Grant Nos. 12575056 and 12205133) and  the Special Fund for Basic Scientific Research of Provincial Universities in Liaoning under grant NO. LS2024Q002.	
\end{acknowledgments}



\begin{thebibliography}{99}
\bibitem{L1}
J. Akin, Y. Zhao, P. G. Kwiat, E. A. Goldschmidt, and K. Fang, Faithful Quantum Teleportation via a Nanophotonic Nonlinear Bell State Analyzer, Phys. Rev. Lett. {\bf134}, 160802 (2025).

\bibitem{L2}
J. Zhao, $et$ $al.$, Enhancing quantum teleportation efficacy with noiseless linear amplification,  Nat. Commun. {\bf14}, 4745  (2023).

\bibitem{L3}
J. Grebel, $et$ $al.$, Bidirectional Multiphoton Communication between Remote Superconducting Nodes, Phys. Rev. Lett. {\bf132}, 047001 (2024).

\bibitem{L4}
A. Z. Ding, $et$ $al.$, Quantum Control of an Oscillator with a Kerr-cat Qubit, Nat. Commun. {\bf16}, 5279 (2025).

\bibitem{L5}
C. Zhang, $et$ $al.$, Experimental Side-Channel-Secure Quantum Key Distribution, Phys. Rev. Lett. {\bf128}, 190503 (2022).


\bibitem{L7}
N. Gisin, G. Ribordy, W. Tittel, and H. Zbinden, Quantum cryptography, Rev. Mod. Phys. {\bf74}, 145 (2002).

\bibitem{L8}
Y. Zhao, $et$ $al.$, Direct photo-patterning of halide perovskites toward machine-learning-assisted erasable photonic cryptography,  Nat. Commun. {\bf16}, 3316 (2025).

\bibitem{L9}
R. Horodecki, P. Horodecki, M. Horodecki, and K. Horodecki, Quantum entanglement, Rev. Mod. Phys. {\bf81}, 865 (2009).

\bibitem{L10}
E. Chitambar, G. Gour, Quantum resource theories, Rev. Mod. Phys. {\bf91}, 025001 (2019).

\bibitem{L11}
A. J. Leggett, Macroscopic quantum systems and the quantum theory of measurement, Prog. Theor. Phys. Suppl. {\bf69}, 80 (1980).

\bibitem{L12}
F. Bibak, F. D. Santo, and B. Daki\'{c}, Quantum Coherence in Networks, Phys. Rev. Lett. {\bf133}, 230201 (2024).

\bibitem{L13}
H. L. Shi, S. Ding, Q. K. Wan, X. H. Wang, and W. L. Yang, Entanglement, Coherence, and Extractable Work in Quantum Batteries, Phys. Rev. Lett. {\bf129}, 130602 (2022).


\bibitem{L14}
F. Ahnefeld, T. Theurer, D. Egloff, J. M. Matera, and M. B. Plenio, Coherence as a Resource for Shor's Algorithm, Phys. Rev. Lett. {\bf129}, 120501 (2022).

\bibitem{L15}
Y. Karli, $et$ $al$., Controlling the photon number coherence of solid-state quantum light sources for quantum cryptography, Npj Quantum Inf. {\bf10},  17 (2024).

\bibitem{L16}
A. Yamauchi, S. Fujiwara, N. Kimizuka, $et$ $al$., Modulation of triplet quantum coherence by guest-induced structural changes in a flexible metal-organic framework. Nat. Commun. {\bf15}, 7622 (2024).

\bibitem{L17}
Y. Wang, Y. Hu, J. P. Guo, J. Gao, B. Song, L. Jiang, A physical derivation of high-flux ion transport in biological channel via quantum ion coherence,  Nat. Commun. {\bf15},  7189 (2024).

\bibitem{L18}
A. Streltsov, G. Adesso, and M. B. Plenio, \emph{Colloquium}: Quantum coherence as a resource, Rev. Mod. Phys. {\bf89}, 041003 (2017).

\bibitem{L19}
C. Cepollaro, $et$ $al.$, Sum of Entanglement and Subsystem Coherence Is Invariant under Quantum Reference Frame Transformations,  Phys. Rev. L {\bf135}, 010201 (2025).


\bibitem{L20}
H. J. Kim, S. Lee, Relation between quantum coherence and quantum entanglement in quantum measurements, Phys. Rev. A {\bf106}, 022401 (2022).



\bibitem{L21}
I. Fuentes-Schuller and R. B. Mann, Alice falls into a black hole: Entanglement in non-inertial frames, Phys. Rev. Lett. {\bf95}, 120404 (2005).

\bibitem{L22}
P. M. Alsing, I. Fuentes-Schuller, R. B. Mann and T. E. Tessier, Entanglement of Dirac fields in noninertial frames, Phys. Rev. A {\bf74}, 032326 (2006).

\bibitem{L23}
Q. Pan and J. Jing, Hawking radiation, entanglement, and teleportation in the background of an asymptotically flat static black hole, Phys. Rev. D {\bf78}, 065015 (2008).

\bibitem{L24}
S. M. Wu, X. W. Fan, R. D. Wang, H. Y. Wu, X. L. Huang and H. S. Zeng, Does Hawking effect always degrade fidelity of quantum teleportation in Schwarzschild spacetime?, J. High Energy Phys. {\bf2023}, 232 (2023).

\bibitem{L25}
A. Ali, S. Al-Kuwari, M. Ghominejad, M. T. Rahim, D. Wang and S. Haddadi, Quantum characteristics near event horizons, Phys. Rev. D {\bf110}, 064001 (2024).

\bibitem{L26}
W. Liu, C. Wen, J. Wang, Lorentz violation alleviates gravitationally induced entanglement degradation, J. High Energ. Phys. {\bf2025}, 184 (2025).

\bibitem{L27}
J. K. Basak, D. Giataganas, S. Mondal and W. Y. Wen, Reflected entropy and Markov gap in noninertial frames, Phys. Rev. D {\bf108}, 125009 (2023).

\bibitem{L28}
E. Mart\'{\i}n-Mart\'{\i}nez, L. J. Garay and J. Le\'{o}n, Unveiling quantum entanglement degradation near a Schwarzschild black hole,  Phys. Rev. D {\bf82}, 064006 (2010).

\bibitem{L29}
W. M. Li, S. M. Wu, Bosonic and fermionic coherence of N-partite states in the background of a dilaton black hole, J. High Energ. Phys. {\bf2024},  144 (2024).

\bibitem{L30}
S. Elghaayda, X. Zhou, M. Mansour, Distribution of distance-based quantum resources outside a radiating Schwarzschild black hole, Class. Quantum Grav. {\bf41}, 195010 (2024).

\bibitem{L31}
S. Sen, A. Mukherjee and S. Gangopadhyay, Entanglement degradation as a tool to detect signatures of modified gravity, Phys. Rev. D {\bf109}, 046012 (2024).

\bibitem{L32}
S. Banerjee, A. K. Alok, S. Omkar and R. Srikanth, Characterization of Unruh channel in the context of open quantum systems, J. High Energy Phys. {\bf2017}, 82 (2017).

\bibitem{L33}
S. Elghaayda, A. Ali, M. Y. Abd-Rabbou, M. Mansour, S. Al-Kuwari,  Quantum correlations and metrological advantage among Unruh-DeWitt detectors in de Sitter spacetime, Eur. Phys. J. C {\bf85}, 447 (2025).

\bibitem{L34}
S. M. Wu, C. X. Wang, D. D. Liu, X. L. Huang, H. S. Zeng, Would quantum coherence be increased by curvature effect in de Sitter space?, J. High Energ Phys. {\bf2023}, 115  (2023).

\bibitem{L35}
S. Harikrishnan, S. Jambulingam, P. P. Rohde and C. Radhakrishnan, Accessible and inaccessible quantum coherence in relativistic quantum systems, Phys. Rev. A {\bf105}, 052403 (2022).


\bibitem{L36}
S. M. Wu and H. S. Zeng, Genuine tripartite nonlocality and entanglement in curved spacetime, Eur. Phys. J. C {\bf82}, 4 (2022).

\bibitem{L37}
H. Dolatkhah, A. Czerwinski, A. Ali, S. Al-Kuwari and S. Haddadi, Tripartite measurement uncertainty in Schwarzschild space-time, Eur. Phys. J. C \textbf{84}, 1162 (2024).

\bibitem{L38}
S. Haddadi, M. A. Yurischev, M. Y. Abd-Rabbou, M. Azizi, M. R. Pourkarimi and M. Ghominejad, Quantumness near a Schwarzschild black hole, Eur. Phys. J. C \textbf{84}, 42 (2024).

\bibitem{L39}
S. M. Wu, X. W. Teng, J. X. Li, S. H. Li, T. H. Liu and J. C. Wang, Genuinely accessible and inaccessible entanglement in Schwarzschild black hole, Phys. Lett. B {\bf848}, 138334 (2024).

\bibitem{L40}
M. M. Du, H. W. Li, S. T. Shen, X. J. Yan, X. Y. Li, L. Zhou, W. Zhong, and Y. B. Sheng, Maximal steered coherence in the background of Schwarzschild space-time, Eur. Phys. J. C {\bf84},
450 (2024).

\bibitem{L41}
A. Chakraborty,  L. Hackl,  M. Zych, Entanglement harvesting in quantum superposed spacetime, Phys. Rev. D {\bf111}, 104052 (2025).



\bibitem{L42}
 S. Barman, I. Chakraborty, S. Mukherjee, Entanglement harvesting for different gravitational wave burst profiles with and without memory, J. High Energy Phys. {\bf2023}, 180 (2023).

\bibitem{L43}
X. L. Huang, X. Y. Jiang, Y. X. Wang, S. Y. Liu, Z. Wang, S. M. Wu, Can boundary configuration be
tuned to optimize directional quantum steering harvesting?, J. High Energ. Phys. {\bf2025}, 23 (2025).

\bibitem{L44}
M. Parikh, F. Wilczek, and G. Zahariade, Quantum Mechanics of Gravitational Waves, Phys. Rev. Lett. {\bf127}, 081602 (2021).


\bibitem{L45}
S. Barman, I. Chakraborty, S. Mukherjee, Signatures of gravitational wave memory in the radiative process of entangled quantum probes, Phys. Rev. D {\bf111}, 025021 (2025).

\bibitem{L46}
Y. Tang, W. Liu, J. Wang, Observational signature of Lorentz violation in acceleration radiation, arXiv:2502.03043.

\bibitem{L47}
L. J. Li, X. K. Song, L. Ye, and D. Wang, Quantifying quantumness in (A)dS spacetimes with Unruh-DeWitt detector, Phys. Rev. D {\bf111}, 065007 (2025).

\bibitem{L48}
Y. K. Zhang, L. J. Li, X. K. Song, L. Ye, D. Wang, Entropic uncertainty and quantum non-classicality of Unruh-Dewitt detectors in relativity, Phys. Lett. B {\bf858},  139063 (2024).

\bibitem{L49}
S. H.  Li, S. H. Shang, S. M. Wu, Does acceleration always degrade quantum entanglement for tetrapartite Unruh-DeWitt detectors?, J. High Energ. Phys. {\bf2025}, 214 (2025).

\bibitem{L50}
S. M. Wu, R. D. Wang, X. L. Huang, Z. Wang, Does gravitational wave assist vacuum steering and Bell nonlocality?, J. High Energy Phys.  {\bf2024},  155 (2024).

\bibitem{L51}
T. Gonzalez-Raya, S. Pirandola and M. Sanz, Satellite-based entanglement distribution and quantum teleportation with continuous variables, Commun. Phys. {\bf7}, 126 (2024).

\bibitem{L52}
W. Izquierdo, J. Beltran, E. Arias, Enhancement of harvesting vacuum entanglement in Cosmic String Spacetime, J. High. Phys. {\bf2025}, 049 (2025).

\bibitem{L53}
R. Li, Z. Zhao, Entanglement harvesting of circularly accelerated detectors with a reflecting boundary, J. High. Phys. {\bf2025}, 185 (2025).

\bibitem{L54}
Y. Ji, J. Zhang, H. Yu, Entanglement harvesting in cosmic string spacetime, J. High. Phys. {\bf2024}, 161 (2024).

\bibitem{L55}
S. Zhang, L. J. Li, X. K. Song, L. Ye, D. Wang, Entanglement and entropy uncertainty in black hole quantum atmosphere, Phys. Lett. B {\bf868}, 139648 (2025).

\bibitem{L56}
Z. Tian, X. Liu, J. Wang, J. Jing, Dissipative dynamics of an open quantum battery in the BTZ spacetime,  J. High Energ. Phys. {\bf2025}, 188 (2025).

\bibitem{L57}
Q. Liu, T. Liu, C. Wen, and J. Wang, Optimal quantum strategy for locating Unruh channels, Phys. Rev. A {\bf110}, 022428 (2024).

\bibitem{L58}
S. Barman, D. Barman, B. R. Majhi, Entanglement harvesting from conformal vacuums between two Unruh-DeWitt detectors moving along null paths, J. High. Energ. Phys. {\bf2022} 106 (2022).

\bibitem{L59}
X. Liu, W. Liu, Z. Liu, J. Wang, Harvesting correlations from BTZ black hole coupled to a Lorentz-violating vector field, J. High Energ. Phys. {\bf2025}  094 (2025).

\bibitem{L60}
S. Kryhin, V. Sudhir, Distinguishable Consequence of Classical Gravity on Quantum Matter,
Phys. Rev. Lett. {\bf134}, 061501  (2025).

\bibitem{LL21}
Y. Tang, W. Liu, Z. Liu, J. Wang, Can the latent signatures of quantum superposition be detected through correlation harvesting?, arXiv:2508.00292

\bibitem{AFL1}
W. Liu, D. Wu, J. Wang, Light rings and shadows of static black holes in effective quantum gravity,
Phys. Lett. B {\bf858}, 139052 (2024).

\bibitem{AFL2}
W. Liu, D. Wu, J. Wang, Light rings and shadows of static black holes in effective quantum gravity II:
A new solution without Cauchy horizons, Phys. Lett. B {\bf868}, 139742 (2025).



\bibitem{L61}
Y. H. Shi, $et$ $al$., Quantum simulation of Hawking radiation and curved spacetime with a superconducting on-chip black hole,  Nat Commun {\bf14}, {\bf3263} (2023).

\bibitem{L62}
Z. Tian, J. Jing, and A. Dragan, Analog cosmological particle generation in a superconducting circuit, Phys. Rev. D {\bf95}, 125003 (2017).

\bibitem{L63}
J. Steinhauer, M. Abuzarli, T. Aladjidi, T. Bienaim\'{e}, C. Piekarski, W. Liu,  E. Giacobino,  A. Bramati, Q. Glorieux, Analogue cosmological particle creation in an ultracold quantum fluid of light, Nat Commun {\bf13}, 2890 (2022).

\bibitem{L64}
J. Steinhauer, Observation of quantum Hawking radiation and its entanglement in an analogue black hole, Nat. Phys.  {\bf12}, 959 (2016).

\bibitem{L65}
Z. Liu, R. Q. Yang, H. Fan, J. Wang, Simulation of the massless Dirac field in 1+1D curved spacetime, Sci. China Phys. Mech. Astron. {\bf68}, 290411 (2025).

\bibitem{L66}
A. J. Brady, I. Agullo,  D. Kranas, Symplectic circuits, entanglement, and stimulated Hawking radiation in analogue gravity, Phys. Rev. D {\bf106}, 105021 (2022).

\bibitem{L67}
I. Agullo, A. J. Brady,  D. Kranas, Quantum Aspects of Stimulated Hawking Radiation in an Optical Analog White-Black Hole Pair, Phys. Rev. Lett. {\bf128}, 091301 (2022).

\bibitem{L68}
Y. Li, $et$ $al$., Microsatellite-based real-time quantum key distribution, Nature  {\bf640}, 47 (2025).

\bibitem{L69}
H. N. Wu, $et$ $al$., Single-Photon Interference over 8.4 km Urban Atmosphere: Toward Testing
Quantum Effects in Curved Spacetime with Photons,  Phys. Rev. Lett. {\bf133}, 020201 (2024).

\bibitem{L70}
 P. Xu, $et$ $al$., Satellite testing of a gravitationally induced quantum decoherence model, Science {\bf366}, 132 (2019).


\bibitem{AD1}
S. B. Giddings, The deepest problem: some perspectives on quantum gravity, 	arXiv:2202.08292

\bibitem{AD2}
G. 't Hooft, Black hole unitarity and antipodal entanglement, Found. Phys. {\bf49}, 1185 (2016).

\bibitem{AD3}
P. Betzios, N. Gaddam, O. Papadoulaki, The Black Hole S-Matrix from Quantum Mechanics,
J. High. Energ. Phys. {\bf2016}, 131 (2016).


\bibitem{AD4}
K. Sravan Kumar, J. Marto, Towards a unitary formulation of quantum field theory in curved space-time: the case of Schwarzschild black hole, Prog. Theor. Exp. Phys.   {\bf 2024}, 123E01 (2024).



\bibitem{L71}
T. Damoar and R. Ruffini, Black-hole evaporation in the Klein-Sauter-Heisenberg-Euler formalism, Phys. Rev. D {\bf14}, 332 (1976).

\bibitem{L72}
S. M. Wu, H. Y. Wu, Y. X. Wang, J. Wang, Gaussian tripartite steering in Schwarzschild black hole,
 Phys. Lett. B {\bf865}, 139493  (2025).

\bibitem{L73}
M. B. Plenio, Logarithmic Negativity: A Full Entanglement Monotone That is not Convex, Phys. Rev. Lett. {\bf95}, 090503 (2005).

\bibitem{L74}
M. A. Nielsen and I. L. Chuang, Quantum Computation and Quantum Information (Cambridge University, Cambridge, 2000).

\bibitem{L75}
T. Baumgratz, M. Cramer, and M. B. Plenio, Quantifying coherence, Phys. Rev. Lett. {\bf113}, 140401 (2014).


\bibitem{LAD1}
X. L. Zong, H. H. Yin, W. Song, Z. L. Cao, Monogamy of Quantum Entanglement, 	Front. Phys. {\bf10}, 880560 (2022).



\end{thebibliography}
\end{document}